\documentclass[%
 reprint,
 amsmath,amssymb,
 aps,
]{revtex4-1}

\usepackage{graphicx}% Include figure files
\usepackage{dcolumn}% Align table columns on decimal point
\usepackage{bm}% bold math
\begin{document}

\preprint{}

\title{
New method of galactic axion search
}% Force line breaks with \\

\author{M. Yoshimura}
\thanks{yoshim@okayama-u.ac.jp}%

\author{N. Sasao}%
 \email{sasao@okayama-u.ac.jp}
\affiliation{%
Research Institute for Interdisciplinary Science,
Okayama University \\
Tsushima-naka 3-1-1 Kita-ku Okayama
700-8530 Japan
%\\
}%

%\affiliation{
%\affiliation{
%\affiliation{%


\date{\today}% It is always \today, today,

\begin{abstract}
An  appealing candidate of the galactic dark matter is the axion,
which was postulated to solve the strong CP (Charge-conjugation Parity)
violation problem in  the standard particle theory.
A new experimental method %under a nearly background-free circumstance
 is proposed to determine  the axion mass.
The method  uses collectively and coherently excited atoms or molecules, 
the trigger laser inducing  galactic axion absorption along with signal photon emission to be detected.

%\begin{description}
%\item[PACS numbers] 14.80.Va, 95.35.+d, 42.50.Nn


{\bf Keywords} \hspace{0.3cm}
   Cold dark matter,
Axion, 
Strong CP problem,
Physics beyond standard model,
Super-radiance

%\end{description}
\end{abstract}

\maketitle

%\tableofcontents

\paragraph*{\bf Introduction}\hspace{0mm}
The galactic dark matter problem may be solved by
 the QCD  (Quantum Chromo-Dynamics) axion \cite{axion 1},  \cite{axion 2},
\cite{cosmic axion},
which provides simultaneously a solution to the conundrum of the
 strong CP problem \cite{pq} in the standard particle theory.
The necessary galactic dark matter mass density
is $\rho_G \simeq 
(0.3 \sim 0.45) \,$GeV/c$^2$\,cm$^{-3}$
which implies by far the largest ambient number density $n_a \sim 10^{13}$cm$^{-3}$ (its precise value
depending on the Peceei-Quinn (PQ) symmetry breaking scale $f_a$).
The current parameter region explored in major experiments is in the range,
$f_a = 10^8 \sim 10^{13}$GeV,
the corresponding axion mass in the range,
$m_a = 10^{2} \sim 10^{-3}$ meV (inversely proportional to $f_a$).

The cosmological origin of the galactic axion is traced back to the QCD
epoch of cosmic temperature $100 \,{\rm MeV}/k_B \sim 10^{12} {\rm K }$,
since the axion mass is generated by the QCD chiral symmetry breaking.
Its couplings
to ordinary quarks, leptons, and gauge bosons are all suppressed by
the PQ symmetry breaking scale $\propto 1/f_a$, hence
axions are born in cold, namely their initial velocities at creation are much less than
the velocity of light.
Cold axions decouple from radiation and matter throughout
the later epoch of cosmological evolution \cite{cosmic axion}.
After the radiation energy density drops below the
axion matter density, gravitational self-interaction
begins to dominate and axions become virialized (process of
violent relaxation \cite{bell}), leaving a
velocity dispersion of order $10^{-3} \times$ the light velocity.

The PQ-symmetry breaking scale may be raised if
it is higher than the reheat temperature
due to inflation \cite{isocurvature perturbation}.
This widens the parameter space to be explored up to $f_a \sim 10^{16}$ GeV.

Ongoing and proposed  experiments \cite{washigton exp},
\cite{capp}, \cite{review of cavity exp},  of galactic axion search 
sensitive to low mass axions use 
magnetic axion conversion into microwave in cavity \cite{haloscope}.
In this paper we propose an entirely new experimental method 
of axion detection using atoms or molecules,
which is sensitive  to the axion mass.
It provides a powerful tool to directly link physics beyond the standard theory and
the dark matter cosmology/astrophysics.

We use the natural unit of $\hbar = c = 1$ unless otherwise stated.\\

\begin{figure*}[htbp]
 \begin{center}
\includegraphics[clip,width=8.0cm]{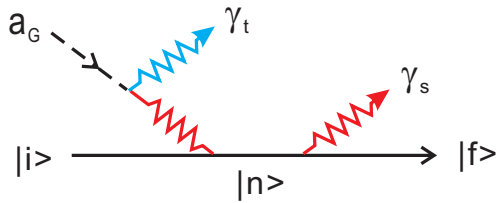} 
   \caption{TRACA diagram of  axion detection.
Virtual or real photon emitted by the  axion $a_G$ decay 
triggered by $\gamma_t$ induces   inelastic photon or Raman
scattering producing a signal photon $\gamma_s$.
Diagram of exchanged atomic/molecular vertexes (s-channel $|n\rangle$ here being replaced by
u-channel $|n\rangle$ not depicted here)
 adds coherently to this contribution.
This diagram is further doubled by $ \gamma_t \leftrightarrow \gamma_s$ exchange.
}
   \label{traca diagram}
 \end{center} 
\end{figure*}

\paragraph*{\bf Microscopic process of galactic axion detection}\hspace{0mm}
Our method proposed here  uses atomic or molecular process,
$a_G +\gamma_t + | i\rangle  \rightarrow | f \rangle  + \gamma_s$:
a galactic axion ($a_G$) collides 
against an atom or a molecule in an excited  state $|i \rangle$,
which then de-excites to a lower energy
state $| f \rangle$, emitting a signal photon  $ \gamma_s$
under the trigger photon $\gamma_t$.
The process is called TRiggered Absorption 
of Cosmic Axion  (TRACA for short).
Relevant diagrams are  depicted in Fig(\ref{traca diagram}). 

The virtual photon of four momentum $k$  may be off or on the mass shell,
depending on its invariant mass squared $k^2= k_0^2 - \vec{k}^2 \neq 0$ or 0.
The two cases have different experimental merits.
In the on-shell case one may view TRACA occurring in two steps of real processes:
the triggered decay axion produces a real photon, which is
inelastically scattered  off excited atom/molecule to
generate a signal photon $\gamma_s$.
The on-shell TRACA may occur in the trigger photon frequency $\omega_t$ at
$ m_a/2$, hence  in the rf (radio frequency) or microwave region, while
the trigger frequency of 
the off-shell TRACA may be in the optical to the infrared (IR) range
depending on the level spacing $\epsilon_{if}$.

We also considered TRACA arising
from diagrams containing direct electron coupling with the axion.
In  most axion parameter region we found
that rates of axion emission are smaller 
than those of Fig(\ref{traca diagram})  even for DFSZ model \cite{axion 2}
of large axion-electron coupling.

The method can  be applied to axion-like particles \cite{alps} as well.

In laboratory experiments on Earth,
 axions collide with its momentum around  $m_a \vec{\beta}_a = -\,m_a \vec{\beta}_e$.
The  axion momentum distribution function $F_a(\vec{q})$ is assumed to have a
momentum dispersion $m_a \beta_0 \simeq 8 \,{\rm neV} (m_a/10 \mu{\rm eV})$,
allowing for a possible BEC component \cite{axion bec}, separately detectable.
The uniform $-\,\vec{\beta}_e$ wind is superimposed by diurnal and sidereal modulation, which
can be detected, too.
One of our ultimate objectives of the proposed experiment 
is to  determine these important parameters.

To maximize TRACA rates, we fully utilize effects of
large axion occupation and coherence (denoted by $\rho_{if}$)
between the initial and final states:
the total TRACA rate for entire atoms is  proportional to
the product of three factors,
(1)  axion  number density $n_a$, giving the enhancement $3 \times 10^{13} m_a/ 10 \mu {\rm eV}$,
(2)   field power $|E_t|^2 = \omega_t n_t$
(its photon number density denoted by $n_t$) driving the axion decay, 
(3) macro-coherently prepared atomic/molecular target number squared   $ N_T^2\rho_{if}^2/V$
with $N_T$ the total number of excited atoms within a volume $V$.
The principle of amplification by macro-coherence \cite{macro-coherence} 
related to the second and the third issues 
has been experimentally demonstrated in \cite{psr exp} for a
weak QED (Quantum ElectroDynamics) process of two-photon emission
in para-hydrogen (p-H$_2$)  vibrational transition (the quantum number change of $v=1 \rightarrow 0$).
The process was amplified  by a factor $\sim 10^{18}$ in rate,
with a realization of atomic coherence $\rho_{if} \sim 0.03$ \cite{ph2 coherence}.
The process was termed PSR (Paired Super-Radiance)
in an obvious analogy to Dicke's super-radiance \cite{sr}.\\

%-------------------------------   figure   ----------------------------------%---------------------   figure   ----------------------------------%

\paragraph*{\bf Derivation of rate formula}\hspace{0mm}
We shall first treat the off mass shell  case
and later discuss briefly the on mass shell case.
Suppose that the intermediate atomic state $|n \rangle $
related by electric dipole transitions to $|i \rangle $ and $|f \rangle $  is far away
in energy from these states; both of
$\epsilon_{ni}\,, \epsilon_{nf} \gg \epsilon_{if}$.
This holds for the example of p-H$_2$.
We further assume the frequency relation, both of $ \omega_t, \omega_s$(trigger and signal
frequencies) $ \gg m_a$ for the off-shell TRACA.

The axion-photon-photon vertex in Fig(\ref{traca diagram}) has 
a form $g_{a\gamma\gamma} a\vec{E}\cdot\vec{B}$ 
in terms of electric $\vec{E}$ and magnetic $\vec{B}$ field operators, since the axion field $a$
is a pseudo-scalar.
Its coupling constant is suppressed by $g_{a\gamma\gamma} \simeq \alpha/(\pi f_a) \,, \alpha = 1/137$.
The dominant atomic couping along the electron line 
is of the usual electric dipole form $\vec{d}\cdot \vec{E}'$,
giving rise to an effective, external field coupling of the  form,
$\vec{B}_t \cdot \langle 0 | T(\vec{E}'  \vec{E}) | 0\rangle \cdot \vec{d}$
with propagation taken into account.
The central part of the probability amplitude is thus
\begin{eqnarray}
&&
\sqrt{ \frac{\rho_G}{2}} \frac{c_{a\gamma\gamma}\,\alpha}{\pi m_a f_a} 
 \left(
\frac{\vec{d}_{nf}\cdot \vec{E}_s  \vec{d}_{ni}\cdot \vec{B}_t }{\epsilon_{ni}}
\frac{(k_t - q)_0^2 }{(k_t- q)^2 }
+ (s \leftrightarrow t )
\right) \times 2
\,.
\nonumber 
\end{eqnarray}
Here $\vec{d}_{nf}, \vec{d}_{ni}$ are dipole operators
and their product divided by the energy denominator   is 
related to $if$ off-diagonal polarizabiity when summed over states $| n \rangle$.
The propagator factor is approximately 
$(k_t - q)_0^2/(k_t- q)^2 \sim - \omega_t/(2 m_a) $
with $q$ the 4-vector of axion.
Model dependent coefficient $ c_{a\gamma\gamma}$ \cite{coupling}, \cite{pdg axion}
is typically $- 0.97$ for KSVZ-model \cite{axion 1},  and  0.36 for DFSZ-model \cite{axion 2}.

For simplicity we take the atomic transition of no spin parity change,
$0^+ \rightarrow 0^+$.
Excluding the squared probability amplitude given above, 
the rate contains a dimensionless quantity integrated over a target volume $V$ prior to taking the square,
\begin{eqnarray}
&&
\hspace*{-0.5cm}
 \int d^3 q |\, \int_V d^3 x\rho_{if} n_T e^{i (\vec{q} - \vec{Q})\cdot \vec{x} }  \,|^2 F_a(\vec{q})
\equiv (\rho_{if} n_T) ^2V^2 A
\,,
\label {geometric factor}
\end{eqnarray}
with $  \vec{Q} = \vec{k}_t + \vec{k}_s - \vec{p}_{if} $ and
$\vec{p}_{if}$  the phase imprinted at the excitation to $|i \rangle $\cite{boosted renp}.
The coherence $\rho_{if}$ and the excited target number density $n_T =N_T/V$
may in principle be a function of target atom positions, but
we take its averaged constant values within the entire target.
We perform the axion momentum integration first, and use the fact that
$q_0 = m_a \beta_0 = 10 \,{\rm neV} (m_a/ 10 \mu{\rm eV})$
much smaller than the inverse of linear size of target volume.
The infinite target volume limit then gives eq.(\ref{geometric factor})
$\sim \rho_{if}^2 n_T N_T (2\pi)^3  \delta (\vec{k}_t + \vec{k}_s - \vec{p}_{if} - \vec{q}_e )\,,
\vec{q}_e = - m_a \vec{\beta}_e$, which cannot be used however.

Due to kinematic reasons one however needs to calculate TRACA rates in
the finite target volume, for which we take a cylinder of radius $R$ and length $L$ along excitation lasers.
This gives the geometric factor defined in eq.(\ref{geometric factor}),
\begin{eqnarray}
&&
A = \frac{1}{ (\pi R^2 L )^2} \left( \frac{2 \sin (Q_{\parallel}L ) }{Q_{\parallel}} \right)^2
\left( \frac{ 2\pi R}{ Q_{\perp}} J_1( Q_{\perp}R )  \right)^2
\,,
\end{eqnarray}
with $J_1(x)$ the Bessel function of the first order.
In our parameter range the dimensionless $A$ is of order $10^{-7}$
when we integrate over all solid angles.

The differential off-shell TRACA rate  is thus calculated as
\begin{eqnarray}
&&
\hspace*{-0.3cm}
\frac{d\Gamma_{{\rm off}}}{d\Omega_s} = \frac{\rho_G}{32\pi^4} 
 (\frac{c_{a\gamma\gamma}\alpha }{m_a f_a })^2
\frac{\mu_{if}^2\epsilon_{if}^2 }{m_a^2 } 
\omega_s^3 E_t^2 \rho_{if}^2 N_T^2 \sin^2 \theta_{{\rm pol}}\,  A 
\,,
\label {traca off-shell rate}
\end{eqnarray}
$\mu_{if}$ the atomic or molecular polarizability.
The relation $E_t^2 = \omega_t n_t$ of the trigger field power to
the photon number density $n_t$ and
various level spacings, $\epsilon_{ab} = \epsilon_a - \epsilon_b$  were used.
We assumed that only one dominant upper level $|n \rangle$ contributes, but extension to include
many contributing levels is straightforward.

A striking feature of the off-shell TRACA formula (\ref{traca off-shell rate}) is
the dependence $\sin^2 \theta_{{\rm pol}}$ with
the relative  angle  $\theta_{{\rm pol}}$ between
polarizations of two photons, $\gamma_t$ and  $\gamma_s$. 
The rate is maximal at the perpendicular $\theta_{{\rm pol}} =\pi/2$
and vanishes at the parallel configuration.

We shall make an order of magnitude estimate of on-shell rate.
The easiest way of calculating the on-shell amplitude
is to replace the real propagator of virtual photon by
the imaginary part in the  zero-width approximation, to give 
\begin{eqnarray}
&&
i \frac{\pi}{ 8}m_a \left( \delta ( \omega_t - \frac{m_a}{2} ) +
\delta ( \omega_s - \frac{m_a}{2} )
\right)
\,,
\end{eqnarray}
for $ (k_t- q)_0^2/ (k_t- q)^2 + (t \rightarrow s )$.
In practice, the  squared delta-function in the rate of on-shell amplitude should be
convoluted with the trigger frequency distribution of a finite width $\Delta$.
Moreover, $(2\pi )^2 \times $ the squared delta-function is replaced by
$2\pi T \times $ the delta-function where $T$ is the time of
trigger irradiation.
This time $T$ is equal to the target length/2.
The procedure ends up in replacing the previous $- \epsilon_{if}/( 2m_a)$  by
$ i \frac{\pi}{ 4} m_a \sqrt{L/\Delta} $.
The on-shell contribution thus dominates over the off-shell contribution for
the axion mass above a threshold; $m_a > O( \sqrt{ \epsilon_{if}\sqrt{\Delta/L}})$
if $\omega_t \simeq m_a/2$ or $\omega_s \simeq m_a/2$.
The axion mass dependence  $\propto m_a^2$ in the on-shell formula
thus derived differs from
the off-shell rate $\propto 1/m_a^2$.

\paragraph*{\bf TRACA rates for p-H$_2$ vibrational transition}\hspace{0mm}
A typical experimental scheme we consider uses
two excitation lasers and a trigger field, all of which
may be irradiated near the same  axis direction.
The target is  of a cylindrical shape whose length
and volume are denoted by $L, V$.

Consider para-hydrogen molecule as a high density  target,
its gas density at the liquid nitrogen temperature 77\,K being
$\sim 5 \times 10^{19}$cm$^{-3}$.
Solid p-H$_2$ target is also  promising from a number of reasons;
(1) a larger molecule density like $2.6 \times 10^{22}$cm$^{-3}$,
(2) expected small relaxation caused by phonon emission.
Relevant phonon emission involves the axion-proton coupling 
predominantly of the spin operator.
The spatial extension of axion wave function is of order cm for 10 $\mu$eV
mass, hence the axion field feels a collective body of para-hydrogen molecule.
Since the molecule has a total nuclear spin 0, this coupling, hence the phonon
relaxation, is highly suppressed.
The vibrational transition of electronically ground state, $v=1 \rightarrow 0$
of level spacing 0.52 eV,  may be used.
The electronically excited level $(1s)(2p)$ is far away from the ground vibrational levels, 
separated by $\sim 10\,$eV.

To calculate the signal rate for molecules, we need
to take into account vibrational modes associated with
electronically excited states such as $(1s)(2p)$.
The effect of  Franck-Condon factor related to the molecular potential curve
of this first electronically excited state may be automatically
avoided by using p-H$_2$ off-diagonal polarizability $\mu_{if} \sim 1.43 \times 10^{-24}$cm$^3$
 \cite{polarizability}.

The  off-shell  TRACA rate for solid p-H$_2$
is numerically
\begin{widetext}
\begin{eqnarray}
&&
\frac{d \Gamma_{{\rm off}}}{d\Omega_s}  \sim 5.9 \times 10^5 \,{\rm sec}^{-1}
(\frac{ 10 \mu\,{\rm eV}}{m_a})^2
x_t (1- x_t)^3 \sin^2 \theta_{{\rm pol}}\, A X 
\,,
\label {ph2 off-shell traca rate}
\\ &&
X = 
 c_{a\gamma\gamma}^2   
 ( \frac{n_T }{2.6 \times 10^{22}\, {\rm cm}^{-3} })^2 
 (\frac{n_t }{10^{18}\, {\rm cm}^{-3} }) 
\,(\frac{\rho_{if}}{0.1})^2  
(\frac{V}{ {\rm cm}^3} )^2
\,, \hspace{0.5cm}
x_t = \frac{\omega_t}{ \epsilon_{if}}
\,.
\label {traca rate 2}
\end{eqnarray}
\end{widetext}
The on-shell TRACA rate is obtained by multiplying 
$\sim 2 \times 10^{-8} (m_a/ 10 \mu {\rm eV})^4 (L/{\rm cm})^2 (100
 {\rm MHz}/\Delta)^2$ to the off-shell rate.

\begin{figure*}[htbp]
 \begin{center}
\includegraphics[clip,width=8.0cm]{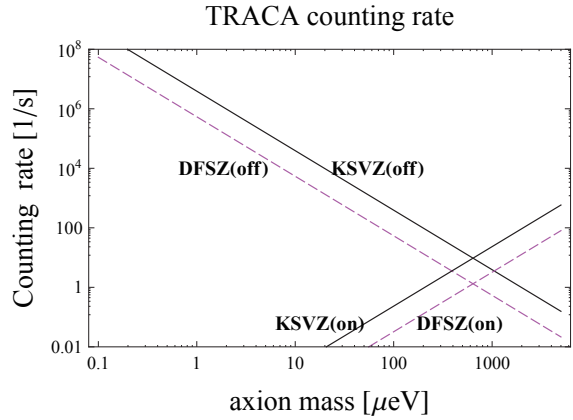} 
   \caption{
TRACA rate for p-H$_2$.
Assumed parameters are the target number density $2.6 \times 10^{22}\, {\rm cm}^{-3}$,
the target volume 1 cm$^3$, the target length 10 cm, the coherence $\rho_{if}=0.1$,
$\omega_t =$ 0.124 eV trigger for the off-shell case
and $\omega_t = m_a/2, \Delta = 300$MHz for the on-shell case
and of a common photon number density $10^{18}\, {\rm cm}^{-3}$ and a common $A = 1$.
We used $c_{a\gamma \gamma}$ values for the KSVZ and DFSZ models,
 and the perpendicular polarization of
the trigger and the signal  was taken, using the formula (\ref{ph2 off-shell traca rate}).
}
   \label{traca rate fig}
 \end{center} 
\end{figure*}

We illustrate total rates of the off-shell and the on-shell TRACA contributions in Fig(\ref{traca rate fig}).
There are higher sensitivities for smaller axion masses less than 100 $\mu$eV.

\paragraph*{\bf Raman excitation and trigger}
\hspace{0cm}
We consider Raman-type of excitation from the ground $v=0$
to $v=1$ state, with frequencies,
 $\omega_1 > \omega_2\,, \omega_1 - \omega_2
= 0.52$eV.
In the on-shell process the trigger frequency may be in the rf or
microwave frequency range,
and the momentum conservation dictates the signal photon emission very near the
excitation axis.
On the other hand, in the off-shell case
it can be in a near infra-red  range.
For example, using CO$_2$ laser of $\omega_t = 0.124$eV for the trigger,
one can relate the angles of trigger irradiation and signal emission as
functions of the axion mass:
$\theta_t = 11 \,{\rm mrad}\sqrt{m_a/10 \mu{\rm eV}}$
and $\theta_s = - 3.5 \,{\rm mrad}\sqrt{m_a/10 \mu{\rm eV}}$
(formulas valid for small axion mass less than  of order 1 meV)
away from both the trigger and the excitation lasers.

\begin{figure*}[htbp]
 \begin{center}
\includegraphics[clip,width=8.0cm]{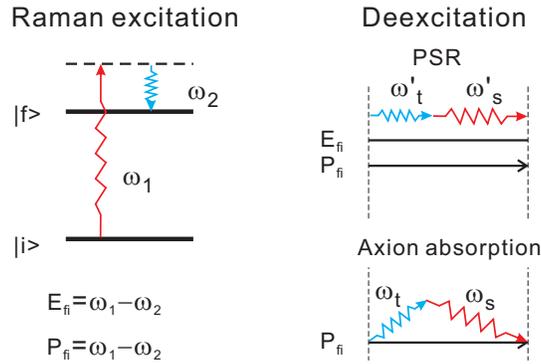} 
   \caption{
Kinematic configuration of initial Raman excitation in the left, the final decay processes in the right.
Both energy and momentum conservations are satisfied 
for PSR only when the trigger ($\omega_t^{\prime}$) is injected
colinearly with the excitation lasers, while they are satisfied for TRACA
when the trigger ($\omega_t$) is injected with angle due to the axion absorption.
These different configurations help separate two processes.
}
   \label{kinetics of excitation and trigger}
 \end{center} 
\end{figure*}

The off-axis configuration and its kinematic constraint is
illustrated in Fig(\ref{kinetics of excitation and trigger}).
This much of the off-axis configuration may be crucial
in rejecting various QED backgrounds (see below on more of this).\\

\paragraph*{\bf Backgrounds}
\hspace{0cm}
By far the largest background may arise from (macro-coherently amplified) PSR process.
The ratio of the signal (off-shell part) to this PSR rate is found to be
$\sim 10^{-27} \left( \frac{10 \mu {\rm eV}}{m_a} \right)^2 \frac{{\cal A}(m_a)}{{\cal A}(m_a=0)} \frac{\cos^2 \theta_{pol}}{\sin^2 \theta_{pol}}$,
taking the same set of target and trigger parameters as indicated in Fig(2).
The last factor comes from the fact polarization vectors of $\gamma_t$ and $\gamma_s$ must be parallel
in the case of the PSR process.
The geometric factor ${\cal A}$ would suppress the background (since ${\cal A}(m_a)\gg {\cal A}(m_a=0)$),
but not entirely due to diffuseness of the finite-size volume effects.
Reduction of the background is also possible by choosing $\theta_{pol}\simeq \pi/2$, but not enough with current technologies.
It is crucial to devise a clever method of PSR background rejection for a realistic experimental proposal.

%\vspace{0.3cm}
In summary,
we proposed a novel experimental method of axion detection using
macro-coherently excited atoms or molecules 
 which can be implemented in small scale laboratories.
The method may clarify how the   galactic axion dark matter is distributed and
serve  to determine the fundamental symmetry breaking scale beyond
the standard particle theory.

\vspace{0.3cm}
{\bf Acknowledgments}
We appreciate discussions with Y. Miyamoto, M. Tanaka, K. Tsumura, and
S. Uetake.
This work is supported in part by JSPS KAKENHI Grant Numbers JP 
15H02093, 15K13486 and 17H02895.
%\end{acknowledgments}

%\appendix


\end{document}